\documentclass[12pt]{iopart}

\begin{document}

\title{What bipartite states are lazy}

\author{Jianwei Xu}

\address{College of Science, Northwest A\&F University, Yangling, Shaanxi 712100, People's Republic of China}
\ead{xxujianwei@nwafu.edu.cn}
\begin{abstract}
A bipartite state is called lazy if the entropy rate of one subsystem is
vanishing for any coupling to the other subsystem. In this paper, we provide
a necessary and sufficient condition for a finite-dimensional bipartite
state to be lazy, and prove that a two-mode Gaussian state is lazy if and only if it is a direct product state.
\end{abstract}


\section{Introduction}

Quantum correlation manifests abundant structures and powerful applications
\cite{Horodecki2009,Modi2012}. How many kinds of quantum correlation and how
to characterize them are therefore quite fundamental questions. Entanglement
and discord, as two kinds of quantum correlation, drawing strong attention,
have been intensively studied, and still in active research (for examples see\cite%
{Chi2013,Liu2013,Spehner2014,Cakmak2013}).

A bipartite state is called lazy, if the entropy rate of one subsystem is
zero for any coupling to the other subsystem. In \cite{Rosario2011}, the
authors established necessary and sufficient conditions for a state to be
lazy. In \cite{Hutter2012}, the authors showed that almost all states are
pretty lazy. It is shown that a maximally entangled pure state is lazy\cite%
{Ferraro2010}, this indicates that the correlation described by lazy states
do not coincide the correlation described by entanglement. In \cite{Xu2014},
by investigating some 2-qubit states, the authors showed that there indeed
exist many lazy states which are entangled, and exist many separable states
which are not lazy.

This paper consider the more general cases. We explore the conditions of a
state to be lazy for arbitrary finite-dimensional bipartite quantum states
and two-mode Gaussian states.

This paper is organized as follows. In Section 2, we establish a necessary
and sufficient condition for a state to be lazy for arbitrary
finite-dimensional bipartite quantum states. In Section 3, we prove that a two-mode Gaussian state is lazy if and only if it is a direct product state. In section 4, we briefly summary this
paper.

As preparations, we briefly review the definition of bipartite lazy state
and introduce some notations. Suppose that quantum systems $A$ and $B$ are
described by the Hilbert spaces $H^{A}$ and $H^{B}$ respectively, the
composite system $AB$ is then described by the Hilbert space $%
H^{AB}=H^{A}\otimes H^{B}$. Let $n_{A}=\dim H^{A}$, $n_{B}=\dim H^{B}$,
being finite or infinite. A state $\rho ^{AB}$ on $H^{AB}$ is called a lazy
state with respect to $A$ if \cite{Rosario2011}
\begin{equation}
C_{A}(\rho ^{AB})=[\rho ^{AB},\rho ^{A}\otimes I^{B}]=0,
\end{equation}%
where $\rho ^{A}=tr_{B}\rho ^{AB}$, $I^{B}$ is the identity operator on $%
H^{B}.$ We often omit $I^{A}$ and $I^{B}$ without any ambiguity. Note that $%
[\rho ^{AB},\rho ^{A}\otimes I^{B}]=0$ keeps invariant under locally unitary
transformations.

An important physical interpretation of lazy states is that the entropy rate
of $A$ is zero in the time evolution under any coupling to $B$ \cite{Rosario2011}
\begin{equation}
C_{A}(\rho ^{AB}(t))=0\Leftrightarrow \frac{d}{dt}tr_{A}[\rho ^{A}(t)\log
_{2}\rho ^{A}(t)]=0.
\end{equation}

\section{lazy states of finite-dimensional bipartite systems}

When $n_{A}=\dim H^{A}$, $n_{B}=\dim H^{B}$ are finite, any state $\rho
^{AB} $ can be expressed as \cite{Schlienz1995}
\begin{equation}
\fl \ \ \ \
\rho ^{AB}=\frac{1}{n_{A}n_{B}}(I^{A}\otimes
I^{B}+\sum_{i=1}^{n_{A}^{2}-1}x_{i}\sigma _{i}\otimes
I^{B}+\sum_{j=1}^{n_{B}^{2}-1}y_{j}I^{A}\otimes \tau
_{j}+\sum_{i=1}^{n_{A}^{2}-1}\sum_{j=1}^{n_{B}^{2}-1}T_{ij}\sigma
_{i}\otimes \tau _{j}).
\end{equation}%
In Eq.(3), we used the $\{\sigma _{i}\}_{i=1}^{n_{A}^{2}-1}$ ($\{\tau
_{j}\}_{j=1}^{n_{B}^{2}-1}$ similarly) defined as
\begin{eqnarray}
\{\sigma _{i}\}_{i=1}^{n_{A}^{2}-1}=\{w_{l},u_{jk},v_{jk}\}, \\
w_{l}=-\sqrt{\frac{2}{l(l+1)}}(P_{11}+P_{22}+...+P_{ll}-lP_{l+1,l+1}), \ \
1\leq l\leq n_{A}-1, \\
u_{jk}=P_{jk}+P_{kj},v_{jk}=i(P_{jk}-P_{kj}), \ \ 1\leq j<k \leq n_{A},
\end{eqnarray}
where $P_{jk}=|j\rangle \langle k|$ with $\{|j\rangle \}_{j=1}^{n_{A}}$ an
orthonormal basis for $H^{A},$ $\{w_{l},u_{jk},v_{jk}\}$ is arranged for any
fixed order. $\{\sigma _{i}\}_{i=1}^{n_{A}^{2}-1}$ are the traceless
generators of $su(n_{A})$ algebra, and fulfill the relations \cite%
{Schlienz1995,Mahler1998}
\begin{equation}
tr\sigma _{i}=0, \ \ \ tr(\sigma _{i}\sigma _{j})=2\delta _{ij},\ \ \
[\sigma _{i},\sigma _{j}]=2i\sum_{k=1}^{n_{A}^{2}-1}f_{ijk}\sigma _{k},
\end{equation}%
where $[\sigma _{i},\sigma _{j}]=\sigma _{i}\sigma _{j}-\sigma _{j}\sigma
_{i},$ $f_{ijk}$ is totally antisymmetric in the subindices $\{ijk\}$. When $%
n_{A}=2$, $\{\sigma _{i}\}_{i=1}^{3}$ are the well known Pauli operators, $%
f_{ijk}$ the permutation symbol.

Now we derive the condition $[\rho ^{AB},\rho ^{A}]=0$ for the state $\rho
^{AB}$ expressed in the form in Eq.(3).
From Eq.(3), we have
\begin{eqnarray}
\rho ^{A}=\frac{1}{n_{A}}(I^{A}+\sum_{i=1}^{n_{A}^{2}-1}x_{i}\sigma
_{i}\otimes I^{B}).
\end{eqnarray}
Then
\begin{eqnarray}
\fl \ \ \ \ \ \ \ \ \ \ \ \
\lbrack \rho ^{AB},\rho ^{A}] =\frac{1}{n_{A}^{2}n_{B}}%
\sum_{ik=1}^{n_{A}^{2}-1}\sum_{j=1}^{n_{B}^{2}-1}[T_{ij}\sigma _{i}\otimes
\tau _{j},x_{k}\sigma _{k}\otimes I^{B}] \ \ \ \ \ \ \ \ \ \ \ \ \ \ \ \ \   \\
\ \ \ \ \ =\frac{1}{n_{A}^{2}n_{B}}\sum_{ik=1}^{n_{A}^{2}-1}%
\sum_{j=1}^{n_{B}^{2}-1}T_{ij}x_{k}[\sigma _{i},\sigma _{k}]\otimes \tau _{j}
\\
\ \ \ \ \ =\frac{2i}{n_{A}^{2}n_{B}}\sum_{ikl=1}^{n_{A}^{2}-1}%
\sum_{j=1}^{n_{B}^{2}-1}T_{ij}x_{k}f_{ikl}\sigma _{l}\otimes \tau _{j}.
\end{eqnarray}
Thus $[\rho ^{AB},\rho ^{A}]=0$ leads to $%
\sum_{ik=1}^{n_{A}^{2}-1}T_{ij}x_{k}f_{ikl}=0$ for any $l$, $j$.

$ $

\textbf{Proposition 1}. $\rho ^{AB}$ in the form in Eq.(3) is lazy respect to $A$ if and only if
\begin{eqnarray}
\sum_{ik=1}^{n_{A}^{2}-1}T_{ij}x_{k}f_{ikl}=0 \ for \  any \ 1\leq l\leq n_{A}^{2}-1,1\leq j\leq n_{B}^{2}-1.
\end{eqnarray}
The lazy states respect to $B$ have the similar result.

$ $

\textbf{Example 1. }As a demonstration, we consider a special class of $%
3\times 3$ states
\begin{eqnarray}
\rho ^{AB}=\frac{1}{9}(I^{A}\otimes I^{B}+\sum_{i=1}^{8}x_{i}\sigma
_{i}\otimes I^{B}+\sum_{j=1}^{8}y_{j}I^{A}\otimes \sigma
_{j}+\sum_{k=1}^{8}\lambda _{k}\sigma _{k}\otimes \sigma _{k}),
\end{eqnarray}
where $\lambda _{k}\neq 0$ for all $k$.

$n_{A}=3,$ then \cite{Mahler1998} $f_{147}=1$, $%
f_{216}=f_{315}=f_{324}=f_{257}=f_{376}=f_{546}=1/2$, $f_{368}=f_{258}=\sqrt{%
3}/2$, notice that $f_{ikl}$ is totally antisymmetric, thus $f_{417}=1,$
etc. Otherwise $f_{ikl}=0.$

For Eq.(13), for any $j$, $l$, Eq.(12) leads to $\sum_{k=1}^{8}\lambda
_{j}x_{k}f_{jkl}=0$, thus $\sum_{k=1}^{8}x_{k}f_{jkl}=F_{jl}=0$. We
explicitly write out the matrix $F=(F_{jl})=0$ as
\begin{equation}
\fl
\left(
\begin{array}{cccccccc}
0 & \frac{x_{6}}{2} & \frac{x_{5}}{2} & -x_{7} & -\frac{x_{3}}{2} & -\frac{%
x_{2}}{2} & x_{4} & 0 \\
-\frac{x_{6}}{2} & 0 & \frac{x_{4}}{2} & -\frac{x_{3}}{2} & -\frac{x_{7}}{2}-%
\frac{\sqrt{3}x_{8}}{2} & \frac{x_{1}}{2} & \frac{x_{5}}{2} & \frac{\sqrt{3}%
x_{5}}{2} \\
-\frac{x_{5}}{2} & -\frac{x_{4}}{2} & 0 & \frac{x_{2}}{2} & \frac{x_{1}}{2}
& \frac{x_{7}}{2}-\frac{\sqrt{3}x_{8}}{2} & -\frac{x_{6}}{2} & \frac{\sqrt{3}%
x_{6}}{2} \\
x_{7} & \frac{x_{3}}{2} & -\frac{x_{2}}{2} & 0 & \frac{x_{6}}{2} & -\frac{%
x_{5}}{2} & -x_{1} & 0 \\
\frac{x_{3}}{2} & \frac{x_{7}}{2}+\frac{\sqrt{3}x_{8}}{2} & -\frac{x_{1}}{2}
& -\frac{x_{6}}{2} & 0 & \frac{x_{4}}{2} & -\frac{x_{2}}{2} & -\frac{\sqrt{3}%
x_{2}}{2} \\
\frac{x_{2}}{2} & -\frac{x_{1}}{2} & -\frac{x_{7}}{2}+\frac{\sqrt{3}x_{8}}{2}
& \frac{x_{5}}{2} & -\frac{x_{4}}{2} & 0 & \frac{x_{3}}{2} & -\frac{\sqrt{3}%
x_{3}}{2} \\
-x_{4} & -\frac{x_{5}}{2} & \frac{x_{6}}{2} & x_{1} & \frac{x_{2}}{2} & -%
\frac{x_{3}}{2} & 0 & 0 \\
0 & -\frac{\sqrt{3}x_{5}}{2} & -\frac{\sqrt{3}x_{6}}{2} & 0 & \frac{\sqrt{3}%
x_{2}}{2} & \frac{\sqrt{3}x_{3}}{2} & 0 & 0%
\end{array}%
\right) =0.
\end{equation}
Consequently, $\rho ^{AB}$ in Eq.(13) with all $\lambda _{k}\neq 0$ is lazy
if and only if $x_{i}=0$ for all $i$.

\section{Two-mode lazy Gaussian states}

Gaussian states are of great practical relevance in quantum information
processing (for recent reviews see \cite{Weedbrook2012,Olivares2012,Wang2007}
etc). The entanglement and discord of two-mode Gaussian states have been
studied \cite{Duan2000,Simon2000,Giorda2010,Adesso2010}. In this section, we
explore that what two-mode Gaussian states are lazy.

Consider system $A$ with continuous variables $\{x_{1},p_{1}\}$ and system $%
B $ with continuous variables $\{x_{2},p_{2}\}$ satisfying $%
[x_{1},p_{1}]=[x_{2},p_{2}]=i$ and $%
[x_{1},x_{2}]=[x_{1},p_{2}]=[p_{1},x_{2}]=[p_{1},p_{2}]=0.$ The creation and
annihilation operators are defined as $a_{j}=(x_{j}+ip_{j})/\sqrt{2}%
,a_{j}^{+}=(x_{j}-ip_{j})/\sqrt{2}$, $j=1,2$. Where + denotes adjoint. The Wigner characteristic
function $\chi (\rho ^{AB},\lambda _{1},\lambda _{2})$ of the two-mode state
$\rho ^{AB}$ is defined as
\begin{eqnarray}
\chi (\rho ^{AB},\lambda _{1},\lambda _{2})=tr(\rho ^{AB}D(\lambda
_{1})D(\lambda _{2})),
\end{eqnarray}
where \
\begin{eqnarray}
D(\lambda _{j})=\exp (\lambda _{j}a_{j}^{+}-\lambda _{j}^{\ast }a_{j})
\end{eqnarray}
is the displacement operator, $\lambda _{j}^{\ast }$ is the complex
conjugate of $\lambda _{j}$.

$\rho ^{AB}$ is Gaussian if $\chi (\rho ^{AB},\lambda _{1},\lambda _{2})$
has the form
\begin{eqnarray}
\chi (\rho ^{AB},\lambda _{1},\lambda _{2}) =\exp[-\frac{1}{2}(\lambda
_{1}^{I},\lambda _{1}^{R},\lambda _{2}^{I},\lambda _{2}^{R})V(\lambda
_{1}^{I},\lambda _{1}^{R},\lambda _{2}^{I},\lambda _{2}^{R})^{t}  \nonumber \\ \ \ \ \ \ \ \ \ \ \ \ \ \ \ \ \ \ \ \ \ \ \ \ \ \ \ \ \ \
-i(d_{1},d_{2},d_{3},d_{4})(\lambda _{1}^{I},\lambda _{1}^{R},\lambda
_{2}^{I},\lambda _{2}^{R})^{t}],
\end{eqnarray}
where the covariance matrix $V$ is a real and symmetric matrix satisfying
the uncertainty relation \cite{Simon1994}, $(d_{1},d_{2},d_{3},d_{4})$ is a
real vector, $\lambda _{j}=\lambda _{j}^{R}+i\lambda _{j}^{I}$.

We now prove Proposition 2 below.

\bigskip

\textbf{Proposition 2.} A two-mode Gaussian state is lazy if and only if it
is a direct product state.

\bigskip

\textbf{Proof.} It is known that up to locally unitary transformations, $\chi (\rho
^{AB},\lambda _{1},\lambda _{2})$ of a Gaussian state $\rho ^{AB}$ can be
written in the form \cite{Duan2000}
\begin{eqnarray}
\chi (\rho ^{AB},\lambda _{1},\lambda _{2}) =\exp[-\frac{1}{2} (\lambda
_{1}^{I},\lambda _{1}^{R},\lambda _{2}^{I},\lambda _{2}^{R})M(\lambda
_{1}^{I},\lambda _{1}^{R},\lambda _{2}^{I},\lambda _{2}^{R})^{t}, \\
M =\left(
\begin{array}{cccc}
n & 0 & c & 0 \\
0 & n & 0 & c^{\prime } \\
c & 0 & m & 0 \\
0 & c^{\prime } & 0 & m%
\end{array}%
\right) ,n\geq 1,m\geq 1.
\end{eqnarray}

Recall the identities (see for example \cite{Cahill1969})
\begin{eqnarray} \ \ \ \ \ \ \ \ \ \
\rho ^{AB} =\int \frac{d^{2}\lambda _{1}}{\pi }\frac{d^{2}\lambda _{2}}{%
\pi }\chi (\rho ^{AB},\lambda _{1},\lambda _{2})D(-\lambda _{1})D(-\lambda
_{2}), \\   \ \ \
\chi (\rho ^{A},\lambda _{1}) =\chi (\rho ^{AB},\lambda _{1},0), \\ \ \ \ \ \ \ \ \ \ \ \ \
\rho ^{A} =\int \frac{d^{2}\lambda _{1}}{\pi }\chi (\rho ^{A},\lambda
_{1})D(-\lambda _{1}), \\
D(\lambda _{1})D(\mu _{1}) =\exp (\frac{\lambda _{1}\mu _{1}^{\ast
}-\lambda _{1}^{\ast }\mu _{1}}{2})D(\lambda _{1}+\mu _{1}).
\end{eqnarray}
Notice that the integrations in this section are all over $(-\infty ,\infty
).$

From Eqs.(20-23), we get%
\begin{eqnarray}
\fl
\lbrack \rho ^{AB},\rho ^{A}] =\int \frac{d^{2}\lambda _{1}}{\pi }\frac{%
d^{2}\lambda _{2}}{\pi }\frac{d^{2}\mu _{1}}{\pi }\chi (\rho ^{AB},\lambda
_{1},\lambda _{2})\chi (\rho ^{AB},\mu _{1})[D(-\lambda _{1})D(-\lambda
_{2}),D(-\mu _{1})] \nonumber \\   \fl   \ \ \ \ \ \ \ \ \ \ \ \
=\int \frac{d^{2}\lambda _{1}}{\pi }\frac{d^{2}\lambda _{2}}{\pi }\frac{%
d^{2}\mu _{1}}{\pi }\chi (\rho ^{AB},\lambda _{1},\lambda _{2})\chi (\rho
^{AB},\mu _{1})[D(-\lambda _{1}),D(-\mu _{1})]D(-\lambda _{2}) \nonumber \\    \fl  \ \ \ \ \ \ \ \ \ \ \ \
=\int \frac{d^{2}\lambda _{1}}{\pi }\frac{d^{2}\lambda _{2}}{\pi }\frac{%
d^{2}\mu _{1}}{\pi }\chi (\rho ^{AB},\lambda _{1},\lambda _{2})\chi (\rho
^{AB},\mu _{1})(e^{\lambda _{1}\mu _{1}^{\ast }-\lambda _{1}^{\ast }\mu
_{1}}-e^{-\lambda _{1}\mu _{1}^{\ast }+\lambda _{1}^{\ast }\mu
_{1}})  \nonumber  \\    \ \ \ \ \ \ \ \ \ \ \ \ \ \ \ \ \ \ \ \ \ \ \ \ \ \ \ \ \ \ \ \ \ \ \ \ \ \ \ \ \ \ \ \ \ \ \ \ \ \ \ \ \
\cdot D(-\lambda _{1}-\mu _{1})D(-\lambda _{2}).
\end{eqnarray}

For any two-mode linear operator $\sigma ^{AB}$, using the Glauber--Sudarshan
$P$ function, $\sigma ^{AB}$ can be expressed as \cite{Glauber1963,Sudarshan1963,Mehta1967}
\begin{eqnarray}
\sigma ^{AB}=\int d^{2}\alpha d^{2}\beta P(\alpha ,\beta )|\alpha
\beta \rangle \langle \alpha \beta |,
\end{eqnarray}
where,
\begin{eqnarray}
\fl  \ \ \ \ \ \ \ \
P(\alpha ,\beta )=\frac{1}{\pi ^{4}}e^{|\alpha |^{2}+|\beta |^{2}}\int
d^{2}ud^{2}v\langle -u,-v|\sigma ^{AB}|uv\rangle e^{u^{\ast }\alpha -u\alpha
^{\ast }}e^{v^{\ast }\beta -v\beta ^{\ast }}e^{|u|^{2}+|v|^{2}},  \\  \fl \ \ \ \ \ \ \ \
\langle -u,-v|\sigma ^{AB}|uv\rangle =e^{-|u|^{2}-|v|^{2}}\int d^{2}\alpha
d^{2}\beta P(\alpha ,\beta )e^{\alpha ^{\ast }u-\alpha u^{\ast }}e^{\beta
^{\ast }v-\beta v^{\ast }}e^{-|\alpha |^{2}-|\beta |^{2}},
\end{eqnarray}
$|\alpha \rangle $,$|u\rangle $ are any coherent states of $A$, $|\beta
\rangle $,$|v\rangle $ are any coherent states of $B$.

From Eqs.(25-27), we see that
\begin{eqnarray}
\fl \ \ \ \ \ \ \ \
\sigma ^{AB}=0\Leftrightarrow P(\alpha ,\beta )=0 \
for \ any \ \alpha ,\beta \Leftrightarrow \langle -u,-v|\sigma ^{AB}|uv\rangle \
for \ any \ u,v.
\end{eqnarray}
Then
\begin{eqnarray}
[\rho ^{AB},\rho ^{A}]=0\Leftrightarrow \langle -u,-v|[\rho ^{AB},\rho
^{A}]|uv\rangle =0 \ for \ any \ u,v.
\end{eqnarray}
From Eq.(24), $\langle -u,-v|[\rho ^{AB},\rho
^{A}]|uv\rangle =0$ reads
\begin{eqnarray}
\int \frac{d^{2}\lambda _{1}}{\pi }\frac{d^{2}\lambda _{2}}{\pi }\frac{%
d^{2}\mu _{1}}{\pi }\chi (\rho ^{AB},\lambda _{1},\lambda _{2})\chi (\rho
^{AB},\mu _{1})(e^{\lambda _{1}\mu _{1}^{\ast }-\lambda _{1}^{\ast }\mu
_{1}}-e^{-\lambda _{1}\mu _{1}^{\ast }+\lambda _{1}^{\ast }\mu _{1}}) \nonumber \\   \ \ \ \ \ \ \ \ \ \ \ \ \ \ \ \ \ \ \ \ \ \ \ \ \ \ \ \ \ \ \ \ \ \ \
\cdot \langle -u|D(-\lambda _{1}-\mu _{1})|u\rangle \langle -v|D(-\lambda _{2})|v\rangle =0.
\end{eqnarray}%
Using the relations (here $\{|i\rangle \}_{i}$ are the number states)
\begin{eqnarray}
|u\rangle =\exp (-\frac{|u|^{2}}{2})\sum_{i=0}^{\infty }|i\rangle
=D(u)|0\rangle ,\ \ \ D(u)^{+}=D(-u), \\
D(u)D(-\lambda _{1}-\mu _{1})D(u) =D(2u-\lambda _{1}-\mu _{1}),
\end{eqnarray}%
and the counterparts for system $B$, Eq.(30) becomes
\begin{eqnarray}
\int \frac{d^{2}\lambda _{1}}{\pi }\frac{d^{2}\lambda _{2}}{\pi }\frac{%
d^{2}\mu _{1}}{\pi }\chi (\rho ^{AB},\lambda _{1},\lambda _{2})\chi (\rho
^{AB},\mu _{1})(e^{\lambda _{1}\mu _{1}^{\ast }-\lambda _{1}^{\ast }\mu
_{1}}-e^{-\lambda _{1}\mu _{1}^{\ast }+\lambda _{1}^{\ast }\mu _{1}})  \nonumber \\ \ \ \ \ \ \ \ \ \ \ \ \ \ \ \ \ \ \ \ \ \ \ \ \ \ \ \ \ \ \ \ \ \ \
\cdot exp[-\frac{|2u-\lambda _{1}-\mu _{1}|^{2}}{2}-\frac{|2v-\lambda _{2}|^{2}}{2}]=0.
\end{eqnarray}
Inserting $\lambda _{j}=\lambda _{j}^{R}+i\lambda _{j}^{I}$, $\mu _{1}=\mu
_{1}^{R}+i\mu _{1}^{I}$ ,$u=u^{R}+iu^{I},v=v^{R}+iv^{I}$ into Eq.(33), we get
\begin{eqnarray}
\ \ \ \
\int \frac{d^{2}\lambda _{1}}{\pi }\frac{d^{2}\lambda _{2}}{\pi }\frac{%
d^{2}\mu _{1}}{\pi }\exp [-\frac{1}{2}\overrightarrow{X}^{t}A_{1}%
\overrightarrow{X}+2\overrightarrow{B}^{t}\overrightarrow{X}] \nonumber \\
=\int \frac{%
d^{2}\lambda _{1}}{\pi }\frac{d^{2}\lambda _{2}}{\pi }\frac{d^{2}\mu _{1}}{%
\pi }\exp [-\frac{1}{2}\overrightarrow{X}^{t}A_{2}\overrightarrow{X}+2%
\overrightarrow{B}^{t}\overrightarrow{X}],
\end{eqnarray}
where
\begin{eqnarray}
\overrightarrow{X}=(\lambda _{1}^{I},\lambda _{1}^{R},\lambda
_{2}^{I},\lambda _{2}^{R},\mu _{1}^{I},\mu _{1}^{R})^{t},\overrightarrow{B}%
=(u^{I},u^{R},v^{I},v^{R},u^{I},u^{R})^{t},
\end{eqnarray}
\begin{eqnarray}
A_{1}=\left(
\begin{array}{cccccc}
2n+1 & 0 & c & 0 & 1 & 2i \\
0 & 2n+1 & 0 & c^{\prime } & -2i & 1 \\
c & 0 & m+1 & 0 & 0 & 0 \\
0 & c^{\prime } & 0 & m+1 & 0 & 0 \\
1 & -2i & 0 & 0 & 1 & 0 \\
2i & 1 & 0 & 0 & 0 & 1%
\end{array}%
\right) ,  \\
A_{2}=\left(
\begin{array}{cccccc}
2n+1 & 0 & c & 0 & 1 & -2i \\
0 & 2n+1 & 0 & c^{\prime } & 2i & 1 \\
c & 0 & m+1 & 0 & 0 & 0 \\
0 & c^{\prime } & 0 & m+1 & 0 & 0 \\
1 & 2i & 0 & 0 & 1 & 0 \\
-2i & 1 & 0 & 0 & 0 & 1%
\end{array}%
\right) ,
\end{eqnarray}
$t$ denotes transpose.

Recall the identity (see for example \cite{Altland2006})
\begin{eqnarray}
\int dz_{1}^{R}dz_{1}^{I}dz_{1}^{R}dz_{1}^{I}...dz_{N}^{R}dz_{N}^{I}\exp [-%
\overrightarrow{Z}^{+}A_{3}\overrightarrow{Z}+\overrightarrow{W}^{+}\overrightarrow{Z}+\overrightarrow{Z}^{+}\overrightarrow{W^{\prime }}]  \nonumber \\
=\pi ^{N}\det (A_{3}^{-1})\exp [\overrightarrow{W}^{+}A^{-1}%
\overrightarrow{W^{\prime }}],
\end{eqnarray}
where $\overrightarrow{Z}%
=(z_{1}^{R}+iz_{1}^{I},...,z_{N}^{R}+iz_{N}^{I})^{t},\overrightarrow{W},%
\overrightarrow{W^{\prime }}$ are two arbitrary complex vectors, $A_{3}$ is a matrix with positive Hermitian part.

It is easy to check that
\begin{eqnarray}
\det (A_{1})=\det (A_{2})=[c^{2}-2(1+m)(2+n)][c^{\prime ^{2}}-2(1+m)(2+n)].
\end{eqnarray}
Using Eqs.(38,39) into Eq.(34), hence Eq.(34) requires that
\begin{eqnarray}
\overrightarrow{B}^{t}A_{1}^{-1}\overrightarrow{B}-\overrightarrow{B}%
^{t}A_{2}^{-1}\overrightarrow{B}=0.
\end{eqnarray}
With direct computation, Eq.(40) reads
\begin{eqnarray}
\frac{8ic^{\prime }}{c^{\prime ^{2}}-2(1+m)(2+n)}u^{I}v^{R}-\frac{8ic}{%
c^{2}-2(1+m)(2+n)}u^{R}v^{I}=0.
\end{eqnarray}
Eq.(41) holds for arbitrary real numbers $u^{I},v^{R},u^{R},v^{I},$ thus
\begin{eqnarray}
c=c^{\prime}=0.
\end{eqnarray}

On the other hand, from Eqs.(15,20-22), it is easy to see that
\begin{eqnarray}
\rho ^{AB}=\rho ^{A}\otimes \rho ^{B}\Leftrightarrow \chi (\rho
^{AB},\lambda _{1},\lambda _{2})=\chi (\rho ^{A},\lambda _{1})\chi (\rho
^{B},\lambda _{2}).
\end{eqnarray}
Together with Eqs.(18,19), we see that
\begin{eqnarray}
c=c^{\prime} =0\Leftrightarrow \rho ^{AB}=\rho ^{A}\otimes \rho ^{B}.
\end{eqnarray}
We then complete this proof.

\section{Summary}

Using the su(n) algebra, we provided a necessary and sufficient condition
for a finite-dimensional bipartite state to be lazy, this condition can be
explicitly checked for a given state in terms of the structure constants $%
\{f_{ijk}\}$ of the su(n) algebra. We also proved that a two-mode Gaussian
states is lazy if and only if it is a direct product state.

How to understand and how to characterize quantum correlation are important
questions in quantum information science. Lazy states possess different
correlation than entanglement and discord, and have an important dynamics
character, i.e., preserving the entropy of subsystem. So the results in this
paper are hopefully interesting for the understandings of quantum
correlation and designing control schemes of quantum systems.

\ack
This work was supported by the National Natural Science Foundation of China
(Grant No.11347213). The author thanks Zi-Qing Wang and Chang-Yong Liu for
helpful discussions.

\end{document}